\documentclass[aps,
prb,
reprint,
preprintnumbers,
superscriptaddress,
amsfonts,
amssymb,
amsmath,
]{revtex4-2}

\usepackage[utf8]{inputenc}
\usepackage{float}
\usepackage{amsmath,bm}
\usepackage[table,x11names]{xcolor}
\usepackage[breaklinks=true,
unicode=true,
urlcolor = RoyalBlue4,
colorlinks = true,
citecolor = RoyalBlue4,
linkcolor = blue
]{hyperref}
\usepackage{graphicx}

\renewcommand{\vec}[1]{\bm{#1}}
\DeclareMathOperator{\sech}{sech}

\begin{document}

\title{Local multiferroic ordering at room temperature in collinear magnetoelectric antiferromagnets induced by flexo-Zeeman coupling}

\author{Paulina~J.~Prusik}
\affiliation{Helmholtz-Zentrum Dresden-Rossendorf e.V., Institute of Ion Beam Physics and Materials Research, 01328 Dresden, Germany}
\affiliation{Faculty of Physics of Dresden University of Technology, 01062 Dresden, Germany}

\author{Igor~Veremchuk}
\affiliation{Helmholtz-Zentrum Dresden-Rossendorf e.V., Institute of Ion Beam Physics and Materials Research, 01328 Dresden, Germany}

\author{Florin~Radu}
\affiliation{Helmholtz-Zentrum Berlin für Materialien und Energie, 12489 Berlin, Germany}

\author{Andrey~N.~Anisimov}
\affiliation{Helmholtz-Zentrum Dresden-Rossendorf e.V., Institute of Ion Beam Physics and Materials Research, 01328 Dresden, Germany}

\author{Pavlo~Makushko}
\affiliation{Helmholtz-Zentrum Dresden-Rossendorf e.V., Institute of Ion Beam Physics and Materials Research, 01328 Dresden, Germany}

\author{Georgy~V.~Astakhov}
\affiliation{Helmholtz-Zentrum Dresden-Rossendorf e.V., Institute of Ion Beam Physics and Materials Research, 01328 Dresden, Germany}

\author{Ren\'e~H\"ubner}
\affiliation{Helmholtz-Zentrum Dresden-Rossendorf e.V., Institute of Ion Beam Physics and Materials Research, 01328 Dresden, Germany}

\author{Alexander~Edstr\"om}
\affiliation{Department of Applied Physics, KTH Royal Institute of Technology, Stockholm, Sweden}

\author{J\"urgen~Fassbender}
\affiliation{Helmholtz-Zentrum Dresden-Rossendorf e.V., Institute of Ion Beam Physics and Materials Research, 01328 Dresden, Germany}
\affiliation{Faculty of Physics of Dresden University of Technology, 01062 Dresden, Germany}

\author{Fabian~Ganss}
\affiliation{Helmholtz-Zentrum Dresden-Rossendorf e.V., Institute of Ion Beam Physics and Materials Research, 01328 Dresden, Germany}

\author{Massimiliano~Stengel}
\affiliation{Institut de Ci\`encia de Materials de Barcelona (ICMAB-CSIC), Campus UAB, 08193 Bellaterra, Spain}
\affiliation{ICREA -- Instituci\'o Catalana de Recerca i Estudis Avan\c{c}ats, 08010 Barcelona, Spain}

\author{Kirill~D.~Belashchenko}
\affiliation{University of Nebraska--Lincoln, Lincoln, NE 68588, USA}

\author{Denys~Makarov}
\email{d.makarov@hzdr.de}
\affiliation{Helmholtz-Zentrum Dresden-Rossendorf e.V., Institute of Ion Beam Physics and Materials Research, 01328 Dresden, Germany}

\author{Oleksandr~V.~Pylypovskyi}
\email{o.pylypovskyi@hzdr.de}
\affiliation{Helmholtz-Zentrum Dresden-Rossendorf e.V., Institute of Ion Beam Physics and Materials Research, 01328 Dresden, Germany}
\affiliation{Kyiv Academic University, 03142 Kyiv, Ukraine}

\date{\today} % Leave empty to omit a date

\begin{abstract}

	Spin-driven multiferroicity attracts significant interest due to its tunability and inherently strong magnetoelectric coupling. While this mechanism induces sizeable electric polarization, it typically occurs at low temperatures and in complex materials. In the simple oxide, the magnetoelectric antiferromagnet  Cr$_2$O$_3$, we experimentally demonstrate the existence of specific domain walls that act as room-temperature multiferroic regions. This behavior stems from an anisotropic crystal-symmetry-dependent mechanism of exchange origin, which is applicable to a broad class of bipartite antiferromagnets. The key signature is the magnetization occurring at antiferromagnetic textures, driven by the flexo-Zeeman interaction. These findings establish a foundation for exploring high-temperature spin-driven multiferroicity for magnetoelectric spin-orbit memory and logic applications.

\end{abstract}

%\keywords{first keyword, second keyword, third keyword}

\maketitle

\section{Introduction}

Materials possessing two and more primary order parameters are known as multiferroics. In particular, the coupling of magnetic and electric degrees of freedom enables the control of magnetic states via electric fields, paving the way to energy-efficient spintronic devices~\cite{Tokura14,Fiebig16,Fert24}. The origin of this so-called magnetoelectric multiferroicity is linked to subtle symmetry tuning that leads to a redistribution of electric charges, depending on the orientation of local magnetic moments and the specific chemical properties of the material~\cite{Hill00}. The available portfolio of such high-temperature, and therefore technologically relevant, materials that also retain their properties as thin films is mainly dominated by the lone-pair antiferromagnetic (AFM) multiferroic BiFeO$_3$ (BFO)~\cite{Wang03a}, doped ferrimagnetic ferrites~\cite{Chun12,Hirose14,Fan25}, and garnets~\cite{Kohara10,Arzamastseva15,Duran17}. Doped BFO is a primary candidate for magnetoelectric spin-orbit (MESO) memory devices~\cite{Husain24,Vaz24,Husain25} yet due to its complexity it still meets challenges for practical applications. This motivates the search for alternative materials that are also AFM-ordered to assure ultrafast dynamics and robustness against external fields. In particular, significant attention is given to spin-driven symmetry lowering in materials where electric dipoles emerge following magnetic ordering~\cite{Baryakhtar84a,Sparavigna94,Mostovoy06,Dzyaloshinskii08,Fiebig16}. Ferroelectricity originating from non-collinear spin ordering naturally provides strong coupling between electric and magnetic degrees of freedom and has been reported for spin spirals~\cite{Kimura03,Goto04,Wang26} and skyrmions~\cite{Seki12a,Ruff15,Hanneken15}. Furthermore, multiferroic domain walls have been observed in rare-earth ferrites~\cite{Hassanpour21,Giraldo21,Zemp25}. These domain walls can act as local conducting paths~\cite{Zhang19a} and can be manipulated by electric fields~\cite{Pyatakov12c,Vakhitov21}. However, even after decades of research, progress regarding spin-driven room-temperature multiferroicity in AFM thin films of chemically simple compounds remains moderate~\cite{Scott13}.

Here, we show that the magnetoelectric simple oxide Cr$_2$O$_3$ supports room-temperature multiferroic domain walls (DWs) of exchange-striction origin at remanence and revealing the largest reported electric polarization for a simple oxide. Specifically, they host experimentally determined polarization and uncompensated magnetization of 0.5\,nC/cm$^2$ and 3\,kA/m, respectively, and are characterized by a non-collinear magnetoelectric coupling coefficient of about $50$\,fC/m quantified by the \textit{ab~initio} calculations. We establish a theoretical foundation for texture-induced magnetization in bipartite AFMs, which we refer to as the flexo-Zeeman interaction. In magnetoelectric AFMs, this leads to an additional non-collinear contribution to the magnetoelectric response, the presence of which we experimentally demonstrate in Cr$_2$O$_3$ with the real-space characterization using scanning nitrogen-vacancy magnetometry (SNVM) and Kelvin probe force microscopy (KPFM). In contrast to the widely studied Cr$_2$O$_3$ DWs parallel to the $c$~axis, this multiferroicity is specific to the DW orientation in the basal plane. The exchange origin of the flexo-Zeeman interaction is confirmed by an analysis of the spin-reorientation transition in an applied magnetic field. Our findings open the possibility of designing spatially anisotropic and highly localized states that simultaneously exhibit multiple ferroic orderings, making them suitable for room-temperature energy efficient (spin)electronics.

\section{Results}

\subsection{Texture-induced magnetization in
antiferromagnets}

\begin{figure*}
	\centering
	\includegraphics[width=\linewidth]{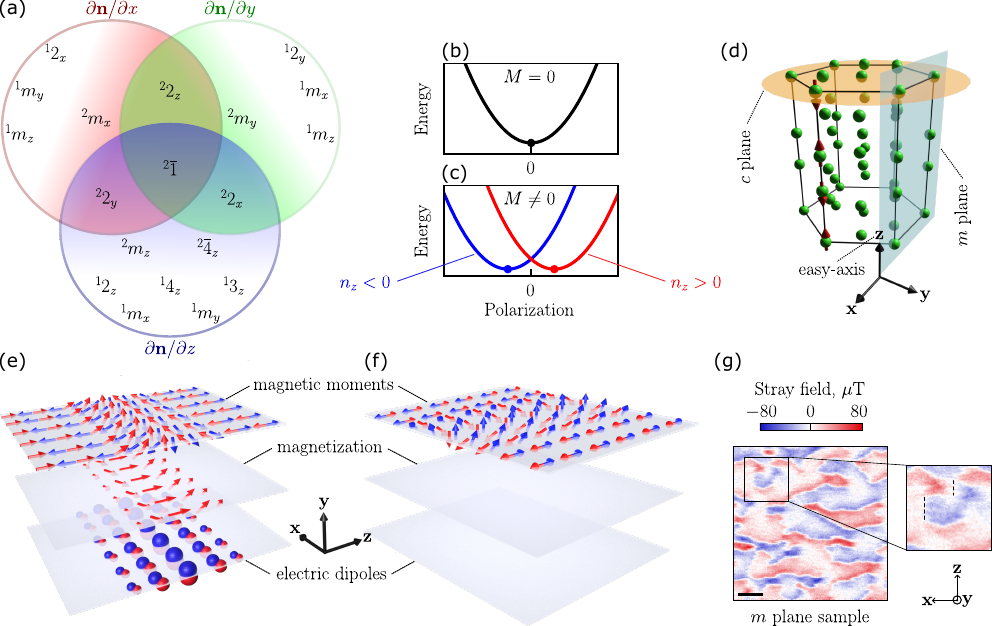}
    \vspace{-0.5cm}
    \caption{\textbf{Flexo-Zeeman interaction in antiferromagnets.} 
    (\textbf{a})~Classification of the symmetry operations within crystallographic point groups that allow for the flexo-Zeeman term with $\partial_i\vec{n}$, $i=x,y,z$. The operations indicated in the white background do not forbid the flexo-Zeeman coupling. 
    (\textbf{b--c})~Energy dependence on polarization: for $\vec{M} \equiv 0$ (\textbf{b}), the minimum is at $P = 0$; otherwise (\textbf{c}), it corresponds to a spatially localized magnetoelectric multiferroic state with finite polarization. (\textbf{d})~Hexagonal cell of chromia that supports flexo-Zeeman coupling $\vec{M} \propto \partial \vec{n}/\partial z$. (\textbf{e},\textbf{f})~Schematic difference between two AFM DWs with different spatial orientation and associated phenomena: the presence or absence of magnetization and electric polarization. Opposite charges are indicated with red and blue spheres. (\textbf{g})~SNVM scan of the $m$-plane thin film of chromia with a stripe stray field pattern. Scale bar is $0.5\,\mu$m. Inset shows two domain walls marked by dashed lines that do not produce stray fields as their planes are parallel to the $c$~axis of the sample.
    }
	\label{fig:intro}
\end{figure*}

It is possible to show that there is a family of bipartite collinear AFMs of specific crystal symmetry, whose \mbox{$\sigma$-model} supports the term %of exchange nature 
that combines an external magnetic field $\vec{H}$ and spatial derivatives of the unit N\'{e}el vector $\vec{n}$. The respective energy density corresponding to this unconventional coupling term is
\begin{equation}\label{eq:pb}
    w_0 = 2b_{ijk} \partial_i n_j H_k,\quad i,j,k=x,y,z,
    % w_0 = 2b_i \partial_i \vec{n} \cdot \vec{H},\quad i=x,y,z,
\end{equation} 
where $b_{ijk}$ are the phenomenological coefficients and Einstein summation notation is used. In particular, it can contain terms of exchange symmetry, $w_0 = 2b_i \partial_i \vec{n} \cdot \vec{H}$ that are generally expected to be stronger than anisotropic relativistic terms. In the following we primarily focus on this high-symmetry case for greater clarity. The macroscopic response tensors involving the N\'{e}el vector are governed by the antisymmetry point groups \cite{Shubnikov51}. We use the standard Litvin notation, in which the left superscript $\sigma = 1,2$ of a symmetry operation $^\sigma\!g$ denotes either preservation ($\sigma = 1$) or interchange ($\sigma = 2$) of the two sublattices. The crystal symmetries of materials supporting this coupling term are summarised in Fig.~\ref{fig:intro}(a). %where the superscript $\sigma=1,2$ of the symmetry operation $^\sigma g$ indicates if the sign of $g\vec{n}$ after reordering of sublattices is kept the same ($\sigma =1$) or not ($\sigma = 2$). 
The necessary condition for the term~\eqref{eq:pb} is that the given $g$ acts on the sign of $\vec{n}$ and the sign of the spatial derivative in the same way, so $\partial_i \vec{n}$ is a common axial vector. In the spirit of naming schemes for interactions in solid and soft matter physics, referring to ``bends'' of magnetic and nematic orderings~\cite{Pyatakov12a}, we refer to the term~\eqref{eq:pb} as the flexo-Zeeman interaction because it combines signatures of a common Zeeman interaction with spatially inhomogeneous AFM textures. On the level of spin ordering, this term enters a continuous version of the spin Hamiltonian as a link between the unitless magnetization vector $\vec{m} = \vec{M}/(2M_\textsc{s})$ and spatial derivatives of the N\'{e}el vector, $w_0' = ({\Lambda}/{M_\textsc{s}}b_i) \vec{m} \cdot \partial_i \vec{n}$, where $M_\textsc{s}$ is the sublattice magnetization, and $\Lambda$ is the constant of the uniform exchange.

The presence of the flexo-Zeeman interaction~\eqref{eq:pb} does not alter a N\'{e}el texture within the interior of a sample exposed to a spatially uniform magnetic field. However, it does influence the surface state for finite-size samples like thin films by modifying the boundary conditions for~$\vec{n}$. Importantly, it is connected with the texture-induced magnetization. In statics, $\vec{m}$ becomes determined not only by the conventional interplay between $\vec{n}$ and $\vec{H}$, but also includes the flexo-Zeeman term
\begin{equation}\label{eq:mag-pb}
    \vec{m} = \vec{n}\times\left[\left( -\frac{1}{M_\textsc{s}} b_i \partial_i\vec{n} +\frac{M_\textsc{s}}{\Lambda} \vec{H} \right)\times \vec{n}\right],\quad i=x,y,z,
\end{equation}
This expression holds for any spatially inhomogeneous distribution of $\vec{n}$ including DWs, skyrmions and spin spirals. In the case of magnetoelectric AFMs, the finite flexo-Zeeman coupling also imposes an additional magnetoelectric response associated with the gradients of $\vec{n}$, which we refer to as the non-collinear magnetoelectric effect. The equilibrium electric polarization~is
\begin{equation}\label{eq:pol-pb}
    P_i = \kappa^e_{ij} E_j + \alpha_{ij}(\vec{n}) H_j + \beta_{ij}(\vec{n}) \partial_j n_i,\quad i,j=x,y,z,
\end{equation}
where $\|\kappa^e\|$ and $\|\alpha\|$ are the tensors of electric and magnetoelectric susceptibilities, $\vec{E}$ is the electric field. The coefficients of the non-collinear magnetoelectric effect~$\beta_{ij}$ are functions of the components of $\vec{n}$ and depend on $b_i$. In particular, $\beta_{ij}$ should vanish with $b_i$ and for a weak magnetoelectric coupling $\beta_{ij} \propto b_i$. As a consequence, it is possible to expect that the respective non-collinear AFM textures will host both, finite magnetization $\vec{M}$~\eqref{eq:mag-pb} and electric polarization $\vec{P}$~\eqref{eq:pol-pb}  at \emph{magnetoelectric} remanence [Fig.~\ref{fig:intro}(b)--(c), Fig.~\ref{fig:intro}(e)--(f)], i.e., demonstrating the behavior characteristic of type-II multiferroics. 

To illustrate the physics associated with the coupling~\eqref{eq:pb}, we focus on the nominally compensated crystallographic cut ($m$~plane) of the collinear AFM $\alpha$-Cr$_2$O$_3$ (chromia) that has the magnetic easy $c$~axis [Fig.~\ref{fig:intro}(d)] and N\'{e}el temperature of $T_\textsc{n} = 308$\,K. Chromia demonstrates room-temperature magnetoelectricity of exchange-striction origin~\cite{Mostovoy10}. Chromia is characterized by the crystal point group $\bar{3}m$ and $(+-+-)$ spin ordering within the unit cell. This leads to the only one non-vanishing coefficient $b:=b_z$ in the exchange approximation. Here, we associate $\vec{\hat{x}}$ and $\vec{\hat{z}}$ axes with the crystallographic $a$ and $c$ directions, respectively [Fig.~\ref{fig:intro}(d)]. Thus, at remanence for chromia, the polarization~\eqref{eq:pol-pb} is
\begin{equation}
    P_i^\text{n-col} = \beta_i(\vec{n}) \partial_z n_i,\quad i=x,y,z
\end{equation}
with $\beta_i := \beta_{iz}$. We note the presence of the contribution of relativistic energy terms in $\vec{P}^\text{n-col}$ not only via $\partial_z\vec{n}$, but also via $\partial_x n_i$ and $\partial_y n_i$.

As it follows from the chromia's symmetry, a possibility to observe stray fields from DWs at the $m$-plane cut of chromia depends on the orientation of the DW. Namely, DWs oriented parallel to the $c$~plane, and referred to as $c$-DWs, should give a pronounced stray field contrast as shown by SNVM imaging in Fig.~\ref{fig:intro}(g). In contrast, DWs parallel to the $c$ axis are not visible in the scans. In the following, we analyze the AFM domain pattern and the associated electric polarization to correlate it with the presence of the flexo-Zeeman term and the local multiferroic ordering. 

\subsection{Multiferroicity at domain walls}

\begin{figure*}[]
	\centering
	\includegraphics[width=\linewidth]{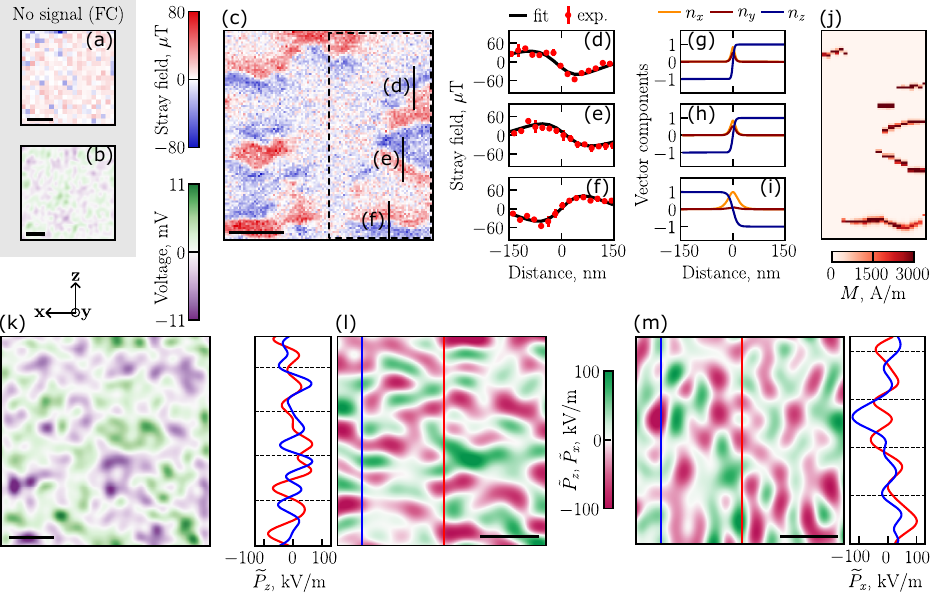}
	\vspace*{-2em}
	\caption{\textbf{Multiferroic response at domain walls in $m$-plane chromia at room temperature.} 
	(\textbf{a})~SNVM and (\textbf{b}) KPFM scans of the $m$-plane thin film after field cooling to set the dominant AFM domain when signatures of domain walls or polarization are absent. Scale bar is $0.5\,\mu$m. FC stands for field cooling.
	(\textbf{c})~SNVM scan of the $m$-plane thin film after zero-field cooling to set a multidomain AFM state. Scale bar is $0.5\,\mu$m. Colorbar is shared with panel~(\textbf{a}). Representative line profiles of DWs shown in panels (\textbf{d--f})~are indicated with vertical solid lines. The corresponding profiles of the N\'{e}el vector $\vec{n}$ are shown in panels (\textbf{G--I}). 
	(\textbf{j})~Magnetization map 
	reconstructed from the region marked by dashed line in panel~(\textbf{c}). 
	(\textbf{k})~KPFM scan of the $m$-plane thin film after zero-field cooling that shows a spatially inhomogeneous electric potential, $\varphi$. Scale bar is $0.5\,\mu$m and the colorbar is shared with panel~(\textbf{b}). 
	(\textbf{l,m})~Components of $\vec{\widetilde{P}}$ proportional to the in-plane electric polarization are recalculated from the data in panel~(\textbf{k}). Side insets show the line scans along blue and red lines, respectively. The dashed lines indicate the major spatial periodicity along the $c$~axis. 
	} 
	\label{fig:scanning}
\end{figure*}

We prepare $m$-plane $120$-nm-thick chromia films and analyze their state using SNVM and KPFM to identify magnetically uncompensated and polar regions over the sample. In this case, the gradient of $\vec{n}$ is parallel to the film surface. First, we perform field cooling through $T_\textsc{n}$ to room temperature to set the preferential AFM domain dominating in the sample~\cite{Pylypovskyi24b}. Over a large-area scan, we do not observe any domain pattern on the sub-$\mu$m scale [Fig.~\ref{fig:scanning}(a)]. KPFM after field cooling shows a comparatively low signal of amplitude of about 1\,mV and weak spatial gradients [Fig.~\ref{fig:scanning}(b)]. 

In contrast to field cooling, after zero-field cooling through $T_\textsc{n}$ to room temperature, a stripe AFM domain pattern in stray fields is observed [Figs.~\ref{fig:intro}(f),~\ref{fig:scanning}(c)], while KPFM contrast increases to 10\,mV with the pronounced spatial gradients [Fig.~\ref{fig:scanning}(k)--(m)]. Fig.~\ref{fig:scanning}(c) is in agreement with the prediction~\eqref{eq:pb} that only those DWs that are nearly perpendicular to the $c$~axis result in finite stray fields because of the emergent magnetization $\vec{M}(\vec{r})$. Their shape and presence is associated with the DW pinning at structural defects like grain boundaries.

Figures~\ref{fig:scanning}(d)--(f) show representative line scans of $c$-DWs that are fitted by the analytical expression according to Eq.~\eqref{eq:mag-pb}, while the reconstructed shape of $c$-DWs in terms of $\vec{n}$ is given in Fig.~\ref{fig:scanning}(g)--(i). The analysis of a straight $c$-DW far from pinning sites gives us the DW width \mbox{$\Delta = \ell = (8 \pm 3)$\,nm}, where $\ell$ is the magnetic length along $c$~axis. This is in agreement with the known reports for the magnetic length in the basal plane~\cite{Hedrich21}. The estimated sublattice magnetization \mbox{$M_\textsc{s}/M_0 = (0.46 \pm 0.12)$} coincides with the literature data $M_\textsc{s}/M_0 = 0.42$~\cite{Hoser12}, where $M_0$ is the sublattice magnetization at~0\,K. This allows us to reconstruct the magnetization distribution across the sample [Fig.~\ref{fig:scanning}(j)] with the amplitude of about 3\,kA/m.
The statistical analysis of the average distance between $c$-DWs reveals the periodicity of \mbox{$(390\pm 40)$\,nm} given by the primary spatial harmonic in the 2D~Fourier transform of the SNVM data [Fig.~\ref{fig:intro}(g),~\ref{fig:scanning}(a)].  

The analysis of the electric state of the sample with AFM domains is given in Fig.~\ref{fig:scanning}(k)--(m). Fig.~\ref{fig:scanning}(k) provides a map of the electric potential $\varphi(x,z)$. The in-plane electric polarization $\vec{P}_{xz}\propto \vec{\widetilde{P}} = \nabla \varphi$ is obtained by computing the Fourier transform and, after filtering the high-frequency noise, is shown in Fig.~\ref{fig:scanning}(l), (m). For $c$-DWs shown in Fig.~\ref{fig:scanning}(g)--(i), the expected polarization components $P_z^\text{n-col} \propto \sech^2 (z/\Delta) \tanh (z/\Delta)$ and $P_x^\text{n-col} = \zeta (n_z\partial_z n_x - n_x \partial_z n_z) \propto \sech (z/\Delta)$, where $\zeta$ is the coefficient of non-collinear magnetoelectric coupling. The amplitude of $\vec{\widetilde{P}}$ allows us to estimate the magnitude of the electric polarization to be about 0.5\,nC/cm$^2$. This number is comparable with the magnetoelectric effect of chromia in the spin-flop phase and corresponds to the coupling between induced magnetization and polarization of about 2\,ps/m. The correlation between the electric and magnetic textures allows us to conclude that the finite polarization appears at $c$-DWs  Additionally, we compare average distances between DWs in SNVM images and characteristic periodicities observed in KPFM. The primary spatial harmonics along the $c$~axis for $P_z$ and $P_x$ give the periodicity $(300 \pm 20)$\,nm and $(350 \pm 40)$\,nm, respectively, which is in a good agreement with the average periodicity of~the~$c$~-~DWs. We further perform \textit{ab~initio} calculations for a Cr$_2$O$_3$ supercell with the non-collinear spin ordering and quantify $\zeta \approx -50$\,fC/m. For a~$c$-DW with $\Delta = 10$\,nm, this gives an electric polarizaiton $P_x^\text{n-col} \approx 0.5$\,nC/cm$^2$, which is close to the experimental estimate.
%\add{. This value is in agreement with the \textit{ab~initio} calculations performed for the Cr$_2$O$_3$ supercell with the non-collinear spin ordering}

\subsection{Quantification of the flexo-Zeeman interaction strength}

\begin{figure*}
	\centering
	\includegraphics[width=1\linewidth]{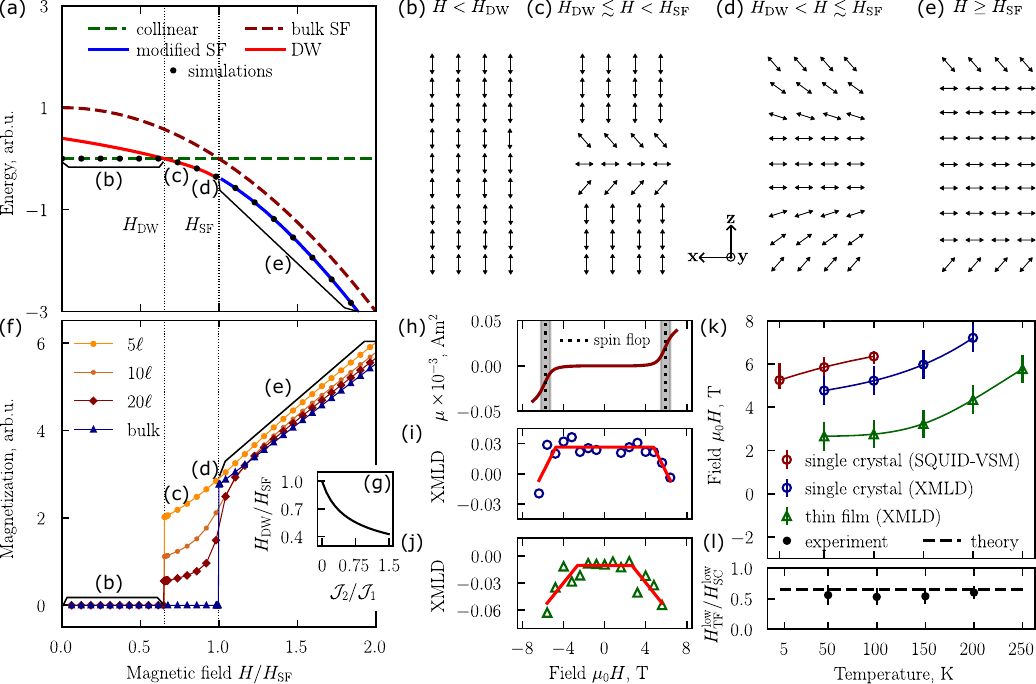}
	%\vspace{-1.5cm}
	\caption{\textbf{Quantification of the strength of the flexo-Zeeman interaction in $c$-plane chromia.} 
	(\textbf{a})~Energy of different magnetic states as a function of the magnetic field for a $10\ell$-thick film. Lines and symbols correspond to the analytics and spin-lattice simulations, respectively.
	(\textbf{b})~Schematics of the collinear state by $\vec{n}$ (shown by double arrows). 
	(\textbf{c})~Schematics of the domain wall state with the gradient along $z$-direction. 
	(\textbf{d})~Schematics of the widening of the domain wall with an increase of magnetic field. 
	(\textbf{e})~Schematics of the spin-flop state in thin film. 
	(\textbf{f})~Simulated magnetization curves as a function of the magnetic field for thin films and single crystal. Field ranges for the magnetic states illustrated in panels (\textbf{b--e}) are indicated. 
	(\textbf{g}) Changes in the ratio of the transition fields $H_\textsc{dw}/H_\textsc{sf}$ as a function of the ratio between the in-plane to out-of-plane exchange coupling strength $\mathcal{J}_2/\mathcal{J}_1$.
	(\textbf{h})~SQUID-VSM measurement of the magnetic moment $\mu$ of a single crystal of chromia at 50\,K. Dotted lines and gray region indicate the spin-flop field and its uncertainty. 
	XMLD signal as a function of magnetic field for (\textbf{i}) a single crystal and (\textbf{j}) 260-nm-thick film measured at $50$\,K. 
	(\textbf{k})~Comparison of the transition fields measured on the single crystal (SQUID-VSM) and estimates of the lower boundaries of critical fields for the spin-reorientation transitions in single crystal and 260-nm-thick film obtained by XMLD. 
	(\textbf{l})~The ratio of the transition fields (lower boundary estimates) at different temperatures for the thin film and single crystal. The dashed line indicates the theoretical expectation of 0.65 [Eq.~\eqref{eq:h-dw-chromia}].}
	\label{fig:spinflop}
\end{figure*}

The coefficient $b$ is of exchange origin and, hence, it can be potentially strong resulting in a substantial impact on the properties of chromia. To quantify $b$, we consider the behavior of $c$-DWs in $c$-plane thin films subjected to a magnetic field parallel to $c$~axis. The spin-flop state in a conventional easy-axis AFM occurs at $H_\textsc{sf} = \sqrt{\Lambda K}/M_\textsc{s}$ with $K$ being the anisotropy constant [green and maroon dashed lines in Fig.~\ref{fig:spinflop}(a), Fig.~\ref{fig:spinflop}(b), (e)]. The presence of a finite magnetic moment on $c$-DWs in chromia due to the flexo-Zeeman interaction allows to lower energy of the film in an applied magnetic field~$H$. Therefore, the DW formation becomes energetically favourable above $H_\textsc{dw} = H_\textsc{sf}/\sqrt{1 + \ell^2 b^2 H_\textsc{sf}^2/A^2 }$, with $A$ being the exchange stiffness along the $c$~axis and $\ell = \sqrt{A/K}$ [red line in Fig.~\ref{fig:spinflop}(a), Fig.~\ref{fig:spinflop}(c)]. In an applied field $H$, the width of $c$-DWs is $\Delta = \ell/\sqrt{1 - (H/H_\textsc{sf})^2}$ [Fig.~\ref{fig:spinflop}(c), (d)].

As a result, the presence of the flexo-Zeeman term leads to the appearance of a two-domain ground state in an AFM exposed to a strong enough magnetic field $H>H_\text{DW}$. In the particular case of chromia, the ratio between $H_\textsc{dw}$ and spin-flop field is
\begin{equation}\label{eq:h-dw-chromia}
    \dfrac{H_\textsc{dw}}{H_\textsc{sf}} = \sqrt{\frac{\mathcal{J}_1}{\mathcal{J}_1+3\mathcal{J}_2}}\approx 0.65,
\end{equation}
where two strongest exchange couplings, the nearest-neighbor $\mathcal{J}_1 \approx 7.5$\,meV (along $c$~axis) and the next-neighbor $\mathcal{J}_2 \approx 3.4$\,meV (within $c$~plane), are considered~\cite{Samuelsen70}. With the modification of the exchange coupling constants, e.g., by strain, the $H_\textsc{dw}$ can be tailored in a wide range [Fig.~\ref{fig:spinflop}(g)].

In the vicinity of and above the $H_\textsc{sf}$, the lowest-energy state corresponds to the spin flop within the film interior accompanied with the tilted surface state at the top and bottom film surfaces [Fig.~\ref{fig:spinflop}(e), blue line in Fig.~\ref{fig:spinflop}(a)]. The presence of the surface state, which is exponentially localized within the thickness $\Xi = \ell/\sqrt{(H/H_\textsc{sf})^2 - 1}$, leads to the lowering of energy in comparison with the uniform spin flop state. The appearance of the two-domain ground state for $H_\textsc{dw} \le H < H_\textsc{sf}$ and the modification of the spin-flop texture in the AFM film are in agreement with the spin-lattice simulations [symbols in Fig.~\ref{fig:spinflop}(a)].

Fig.~\ref{fig:spinflop}(f) shows the relative magnetic moment of chromia films of different thickness $L$ in the magnetic field. For large $L$, the relative contribution of the $c$-DW magnetization comparing to the spin-flop state is small [maroon line in Fig.~\ref{fig:spinflop}(f)]. In thinner samples of up to several hundredths of~nm (several dozens of~$\ell$) [dark and bright orange lines in Fig.~\ref{fig:spinflop}(f)], the transition from the $c$-DW to spin-flop spin configurations becomes hardly distinguishable. Hence, the experimental determination of the spin flop transition in thin chromia films (corresponding to $H_\textsc{dw}$) compared to bulk samples (corresponding to $H_\textsc{sf}$) could provide access to the strength of $b$. To maximize the DW contribution, the thickness of thin film should be comparable to the spatial extent occupied by the  $c$-DW. 

The predicted modification of the transition field is accessed experimentally by X-ray magnetic linear dichroism (XMLD) measurements~\cite{Luo19a} of $c$-plane-oriented chromia single crystals and 260-nm-thick films (about $25\ell$ in thickness). The single crystal is additionally measured using a superconductive quantum interference device -- vibrating sample magnetometer (SQUID-VSM) [Fig.~\ref{fig:spinflop}(h)]. The changes of the magnetic state measured by XMLD are clearly visible in both single crystal and thin film samples [Fig.~\ref{fig:spinflop}(i), (j)]. For the single crystal chromia sample, the measured spin-flop field is close to the literature data and agrees with SQUID-VSM. In contrast, the spin-reorientation field of the thin film sample is substantially lower [green line in Fig.~\ref{fig:spinflop}(k)]. The ratio between the transition fields measured for the thin film and single crystal is $0.56\pm0.07$ [Fig.~\ref{fig:spinflop}(l)], which matches the theoretical prediction in Eq.~\eqref{eq:h-dw-chromia}. Considering that in thin film samples the difference between the $c$-DW and spin-flop states is hardly distinguishable [orange line for $5\ell$, Fig.~\ref{fig:spinflop}(f)], we anticipate that the experimentally observed lowering of the spin-reorientation field for the 260-nm-thick sample is in agreement with the presence of the $c$-DW. Having confirmed the presence of the domain-wall state in thin film samples, we can provide an estimate for the flexo-Zeeman coefficient, $b = 37 \,\mu \text{A}$ at low temperatures. 

\section{Discussion}

We provide a theoretical framework and design rules for engineering nanoscale AFM polar regions within non-collinear AFM textures via flexo-Zeeman interaction, with potential applications in memory and logic technologies. Although magnetic fields are often considered irrelevant for AFMs, the flexo-Zeeman interaction enables a~non-collinear magnetoelectric response. We demonstrate this effect experimentally by identifying suitable DWs in chromia and quantify the coupling coefficient via \textit{ab~initio} calculations. In addition to finite magnetization, these antiferromagnetic DWs support electric polarization and hence resemble magnetoelectric multiferroics. The detected electric polarization is the largest at room temperature for simple oxides. We anticipate that this remanent state in chromia and other materials supporting the flexo-Zeeman interaction can also be created globally across the entire film via strain gradients that induce gradients in the N\'{e}el vector~\cite{Makushko22}. This opens up new possibilities for the design of magnetically or electrically reconfigurable~\cite{Belashchenko16}, DW-based electric channels for the MESO concept, as well as for sensorics applications through the pinning of DWs at mesostructures~\cite{Hedrich21} or notches. 

The consequences of the flexo-Zeeman interaction, which is symmetry-allowed in a wide class of bipartite AFMs, expand upon earlier predictions regarding the possibility of inducing polarization~\cite{Stengel24} and texture-driven magnetization~\cite{Papanicolaou95,Andreev96,Pylypovskyi21e,Makushko22}. This concept can be extended to more complex AFM orderings and global non-collinear states, such as spin spirals and even offer possibilities to manipulate antiferromagnetic textures through applied field gradients. The spin-driven mechanism of magnetoelectricity in chromia leads to a sizable electric polarization at the DWs. By applying our findings to other materials, we anticipate identifying even stronger room-temperature multiferroicity of spin-driven origin.

\section*{Acknowledgments}

The authors thank Dr.~Nina Elkina (HZDR and University of D\"{u}sseldorf) and Prof.~Patrick Maletinsky (Basel University) for discussions on numerical approaches for the reconstruction of AFM domain patterns, Prof. Nicola~A.~Spaldin, Dr. Sophie~F.~Weber, and  Prof.~Claude Ederer (ETH Z\"{u}rich) for their valuable input on \textit{ab~initio} calculations of chromia properties, Annette Kunz and Andreas Worbs (HZDR) for TEM specimen preparation, and Dr. Shengqiang Zhou (HZDR) for providing access to the SQUID-VSM measurement device. Numerical calculations are performed using the Hemera high-performance cluster at  HZDR~\cite{hzdrcluster}. Support by the Structural Characterization Facilities Rossendorf at the Ion Beam Center (IBC) at HZDR is greatly appreciated.

\section*{Funding}
This project was financed in part via ERC grant 3DmultiFerro (Project No. 101141331) and German Research Foundation (DFG; project \#VE948/5–1). K.B. acknowledges support from the National Science Foundation through Grant No. OIA-2521415 and from a UNL Grand Challenges catalyst award entitled Quantum Approaches Addressing Global Threats. F.R. acknowledges financial support for the VEKMAG end station at the BESSY-II by the German Federal Ministry for Education and Research (BMBF 05K10PC2, 05K10WR1, 05K10KE1). A.N.A and G.V.A acknowledge financial support from the project Quantum Sensing for Fundamental Physics (QS4Physics) from the Innovation pool of the research field Helmholtz Matter of the Helmholtz Association. M.S. acknowledges support by the Spanish MCIN/AEI/10.13039/501100011033 through grants PID2023-152710NB-I00 and CEX2023-001263-S, as well as by the Generalitat de Catalunya through grant 2021 SGR 01519.

\section*{Author contribution}
O.V.P and D.M. conceived and supervised the project. 
O.V.P. proposed the concept of flexo-Zeeman interaction and developed the analytics for the analysis of the finite-temperature phenomena. 
P.J.P. performed the analytical calculations, simulations, and analysis of SNVM, SQUID-VSM, and KPFM measurements with support from O.V.P, I.V., A.N.A., G.V.A., P.M.,  K.D.B., J.F., and D.M.
I.V., P.M., and F.R. performed the XMLD measurements. 
A.N.A., P.M., and G.V.A. performed the SNVM and KPFM measurements. 
I.V. and P.M. fabricated thin-film samples, performed magnetotransport and SQUID-VSM characterization. 
R.H. performed and interpreted the transmission electron microscopy analyses. 
F.G. and I.V. performed and interpreted the X-ray diffraction measurements. 
P.J.P., O.V.P., A.E., M.S. and K.D.B. performed and discussed the symmetry analysis and correlated material parameters with \textit{ab~initio} calculations. 
O.V.P., P.J.P., and D.M. wrote the manuscript with the input from all authors. 
All authors discussed the results and reviewed the manuscript.

\section*{Competing interests}
There are no competing interests to declare.

% \bibliography{dw-spin-flop}


\end{document}